\documentclass[osajnl,twocolumn,showpacs,floatfix,superscriptaddress,10pt]{revtex4-1} 
\usepackage{amsmath,amssymb,graphicx}
\usepackage{color, amssymb, graphicx, amsfonts }
\usepackage{multirow,amssymb,amsbsy,amsmath}
\usepackage[breaklinks=true,colorlinks=true,linkcolor=blue,urlcolor=blue,citecolor=blue]{hyperref}
\usepackage{array,booktabs}
\usepackage{verbatim}
\usepackage{bm}
\usepackage{bbold}
\usepackage{appendix}
\usepackage{float}
\begin{document}
\title{Territories of Parrondo's paradox and its relation with entanglement in quantum walks}

\author{Munsif Jan}\email{Corresponding author: mjansafi@zjnu.edu.cn}
\author{Niaz Ali Khan}
\author{Gao Xianlong}\email{Corresponding author: gaoxl@zjnu.edu.cn}
\affiliation{ Department of Physics, Zhejiang Normal University, Jinhua 321004, P. R. China}

\begin{abstract}
Parrondo's paradox is a well-known counterintuitive phenomenon, where the combination of unfavorable situations can establish favorable ones. In this paper, we study one-dimensional discrete-time quantum walks, manipulating two different coins (two-state) operators representing two losing games A and B, respectively, to create the Parrondo effect in the quantum domain. We exhibit that games A and B are losing games when played individually but could produce a winning expectation when played alternatively for a particular sequence of different periods for distinct choices of the relative phase. Furthermore, we investigate the regimes of the relative phase of the initial state of coins where Parrondo games exist. Moreover, we also analyze the relationships between Parrondo's game and quantum entanglement and show regimes where the Parrondo sequence may generate a maximal entangler state in our scheme. Along with the applications of different kinds of quantum walks, our outcomes potentially encourage the development of new quantum algorithms.
\end{abstract}



\maketitle 

\section{Introduction}
Parrondo's paradox, which was originally based on flashing Brownian ratchet\,\cite{parrondo1996eec,parrondo2000new}, characterizes counterintuitive gambling games where two individually losing games can construct a winning game when combined it in a right way\,\cite{harmer1999game,harmer1999parrondo}. The original Parrondo's paradox has two versions, referred as capital-dependent\,\cite{harmer1999game,harmer1999parrondo,jian2014beyond} and history-dependent\,\cite{parrondo2000new,parrondo2004brownian}. The only difference between these two versions is just the switching mechanism of game B. These paradoxical mechanism has been observed in many situations of interest, for example, counterintuitive drift in the physics of granular flow\,\cite{PhysRevLett.58.1038}, enzyme transport analyzed by a four-state rate mode\,\cite{westerhoff1986enzymes}, combination of declining branching processes produces an increase\,\cite{key1987computable}, and finance model where capital increases by investing an assets with negative growth rate\,\cite{maslov1998optimal}. In the last two decades, the Parrondo's paradox has been extended to various fields ranging from physics\,\cite{flitney2002quantum,lee2002quantum,PhysRevLett.88.024103,khan2010quantum,ethier2012parrondo,pawela2013cooperative,pejic2015quantum,grunbaum2016maximal}, population genetics\,\cite{masuda2004subcritical,wolf2005diversity,reed2007two}, and even to economics\,\cite{spurgin2005switching}, has attracted a particular attention due to its potential to characterize the strategy of altering the unstable situation into a stable ones. Although the paradox has been proposed theoretically both in the classical and quantum systems\,\cite{flitney2002quantum,grunbaum2016maximal,lee2002quantum,pawela2013cooperative,pejic2015quantum}. It is well-known that the classical Parrondo effect is a type of random walk that can be described by the Fokker-Planck equation\,\cite{Amengual2004},which is a Wick rotation of the Schr\"odinger equation\,\cite{Cannata_1999}, thus, there are deep interconnections between these types of classical random walks and quantum walks. Our work motivates the further exploration of these ideas.

Quantum walks (QWs)\,\cite{PhysRevA.48.1687,venegas2012quantum}, which are natural extensions of the classical random walks (CRWs) in the quantum domain, possessing a quadratic gain over the CRWs\,\cite{mackay2002quantum} due to the remarkable features like interference and superposition. Thus, QWs offer a flexible and powerful platform to investigate different physics, ranging from the design of efficient algorithms in quantum information processing\,\cite{Santha2008,Portugal2013,Childs2014} (even constructing universal quantum computation\,\cite{Childs2009,Childs2013}), the realization of exotic physical phenomena in the context of topological phases\,\cite{PhysRevA.82.033429,Kitagawa2012a,Kitagawa2012b}, to quantum physics out of equilibrium\,\cite{Eisert2015} (for example, observing the dynamical quantum phase transition \cite{PhysRevLett.121.130603,heyl2018dynamical,xu2018measuring} and even investigating quantum thermodynamics\,\cite{Parrondo2015,Garnerone2012,Romanelli2012,Romanelli2014,Romanelli2015}). The Parrondo effect in the quantum walks has been proposed in \,\cite{Flitney2004,chandrashekar2011parrondo,flitney2012quantum,li2013quantum,rajendran2018playing,rajendran2018implementing,Pires2020,Walczak2021}, using different strategies and can also be viewed as a special sample of the disordered QWs\,\cite{PhysRevLett.106.180403,PhysRevA.97.012116}. The periodic sequence and dynamically disordered QWs, which possess some distinctive properties, such as, enhancing the entanglement between the coin and the position\,\cite{PhysRevLett.111.180503,Vieira2014,wang2018dynamic}. Unlike the CRWs, QWs are characterized by quantum superpositions of amplitudes rather than classical probability distributions. However, the coherent character of the QW plays a vital role in the realization of a quantum Parrondo game. Recently, we have experimentally realized the quantum version of the Parrondo effect\,\cite{jan2020} in delayed-choice QW, based on currently developed compact large-scale QW platform\,\cite{Xu2018}. Furthermore, it has been observed that the quantum Parrondo effect is disappeared by decohering QW with a pure dephasing channel, implying that the quantum coherence of QWs plays a vital role in the emergence of the quantum Parrondo effect\,\cite{jan2020}.

Here, we demonstrate the genuine quantum Parrondo effect in one-dimensional (1D) discrete time QWs using two different coin operations, i.e., $C_\text{A}$ and $C_\text{B}$, representing games A and B, respectively. We show that games A and B are individually losing games. However, when we play these two games alternatively in the particular sequence of different periods, could produce a winning game which is known as Parrondo's paradox. Recent attempts\,\cite{li2013quantum,flitney2012quantum,rajendran2018implementing} have been failed to realize the genuine Parrondo game in QWs for the case of a two-state coin (qubit) over the finite and infinite steps. Further, we show that the presence of wait state in the shift operator of QWs is not the necessary condition for the occurrence of genuine Parrondo's paradox. Hence, we have shown a number of sets of coin rotation parameters for which genuine Parrondo's paradox exist for the particular sequence of almost all periods. In addition, we also discuss the relationship between quantum Parrondo's games and coin-position entanglement along with regimes of the relative phase $\eta$, where Parrondo's games exist and may generate maximal entangler state in our considered scenarios.


We suppose 1D discrete-time QWs whose total Hilbert space can be expressed as $H=H_{c}\otimes H_{p}$, where $H_{c}$ is a two-dimensional coin space spanned by \{$|0\rangle,|1\rangle\}$, and $H_{p}$ represents an infinite-dimensional site space spanned by $|x\rangle\,(x\in Z$ is the site). Each step of the QW possesses two operations $C$ and $S$: $C$ is the rotation of the coin state in $H_C$, and $S$ is the shift operator which describes the movement of the walker according to the coin state. Generally, the coin rotation operator of a two-dimensional space is defined by three parameters ($\alpha,\beta,\gamma$), i.e.,
\begin{align}
	C(\alpha,\beta,\gamma)=\left( \begin{array}{cc} 
		e^{i \alpha} \cos\beta & 	-e^{-i \gamma} \sin\beta\\
		e^{i \gamma} \sin\beta & e^{-i \alpha} \cos\beta
	\end{array}\right).
\end{align}
Different rotation operators describe distinct QWs correspond to different games. The shift operator can be defined as 
\begin{align}
		\textit{S}=\sum^{\infty}_{x = -\infty}(|0\rangle \langle 0|\otimes|x+1\rangle_{p}\langle x|_{p} +|1\rangle \langle 1|\otimes|x-1\rangle_{p}\langle x|_{p}).
\end{align}
\begin{figure}[h]
\includegraphics[width=0.9\linewidth]{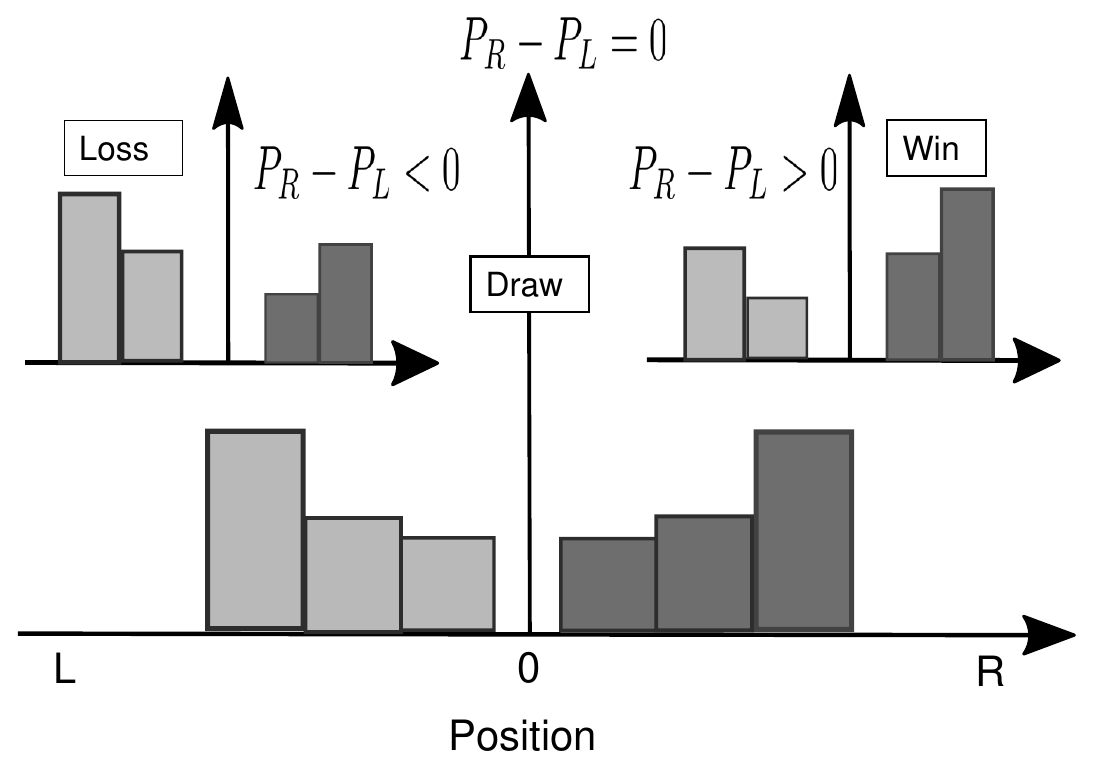}
\caption{(Color online) An illustration of a losing versus a winning strategy in a 1D QW. Grey and black distributions show the probabilities of the walker to the left $P_{\text{L}}$ and right $P_{\text{R}}$ of the origin, respectively.}\label{Fig1}
\end{figure}
Therefore, the single step evolution of the QW can be described as: $U=S{\cdot}C(\alpha,\beta,\gamma)$. In our scheme, the generic unbiased initial state of the walker is prepared as $|\Psi(0) \rangle = \frac{1}{\sqrt{2}} (|0\rangle +e^{i \eta}|1\rangle)_{\text{c}}\otimes|x\rangle_{\text{p}}$, where the subscript $c$ ($p$) represents the coin (position) state, respectively. The global state at time step $t$ ($t$ is an integer) then reads as $|\Psi(t)\rangle=U^t|\Psi_0\rangle$.

To demonstrate the paradoxical scenario of Parrondo game in a 1D discrete-time QW, we define a game on the state $\Psi(t)$. Note that we can use the bias of the probability distribution of $\Psi(t)$ to define the state of winning and losing outcome of the game. The winning and losing of game can be decided according to the strategy shown in Fig.~\ref{Fig1}: if $P_{\text{R}}-P_{\text{L}}>0$ ($P_{\text{L}}=\sum^{-1}_{x=-\infty}|\langle x|\Psi(t)\rangle|^2$ and $P_{\text{R}}=\sum^{\infty}_{x=1}|\langle x|\Psi(t)\rangle|^2$), which means that the walker has a greater probability of appearing at the right of the origin, representing a winning game; on the contradictory, if $P_{\text{R}}-P_{\text{L}}<0$, the game is losing; when $P_{\text{R}}=P_{\text{L}}$, this represents a draw situation. If $P_{\text{R}}-P_{\text{L}} > 0$ is maintained throughout the dynamics, it indicates a winning expectation of the game. Similarly, the converse situation denotes a losing expectation as shown in Fig.~\ref{Fig1}. 
\begin{figure}[htbp]
\centering
	\includegraphics[width=0.9\linewidth]{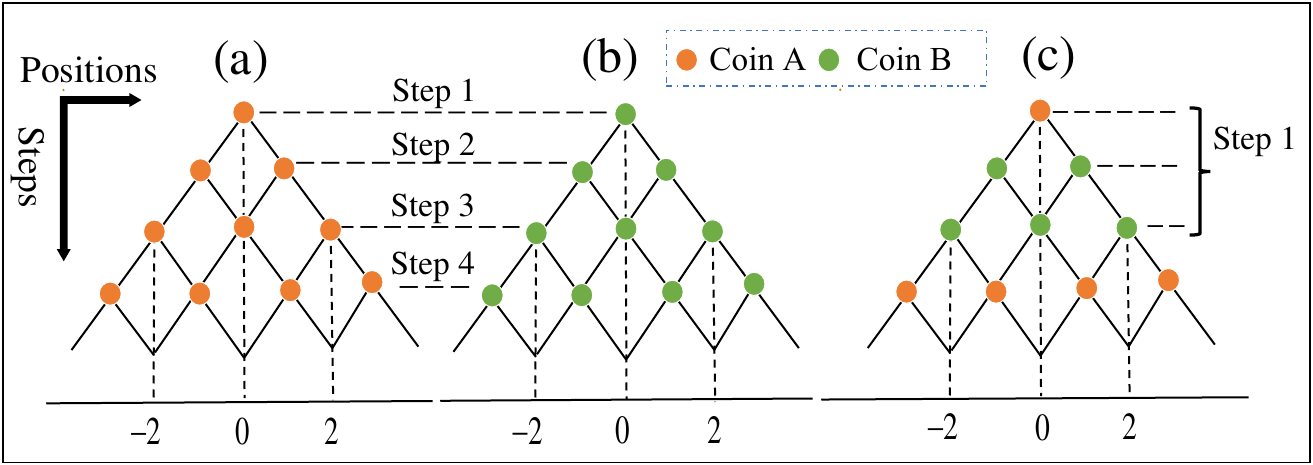}
\caption{(Color online) An illustration of a coin operation in a 1D QW for (a) game A realized with $C_{\text{A}}$ (orange), (b) game B realized with $C_{\text{B}}$ (green), and (c) game ABB played with both coins in the periodic sequence of $C_{\text{A}}C_{\text{B}}C_{\text{B}}$, respectively.} \label{Fig2}
\end{figure}

\begin{figure*}[ht!]
	{\includegraphics[width=1\linewidth]{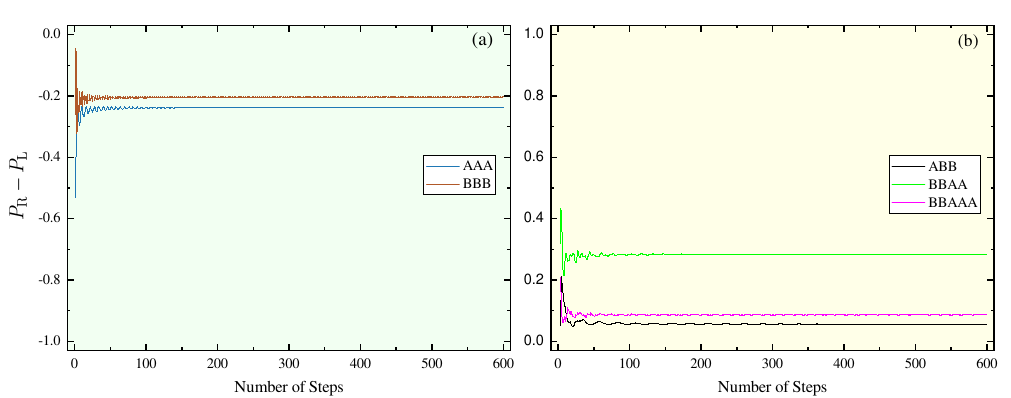}}
    \caption{(color online) Walker difference probability distribution $P_{\text{R}}-P_{\text{L}}$ versus number of steps for (a) game A with coin operation $C_{\text{A}}(150,30,172)$, game B with coin operation $C_{\text{B}}(175,65,165)$, individually. (b) when played alternatively in particular sequences of different periods regarded as different games, e.g., game ABB with coin operation $C_{\text{A}}C_{\text{B}}C_{\text{B}}$, similarly, for BBAA, and BBAAA, with $\eta=3\pi/2$. These figures demonstrate the occurrence of Parrondo's paradox in 1D discrete-time QWs over infinite number of steps for coins operation $C_{\text{A}}(150,30,172)$ and $C_{\text{B}}(175,65,165)$.}\label{Fig3}
\end{figure*}

\begin{figure*}[hbtp!]
	{\includegraphics[width=1\linewidth]{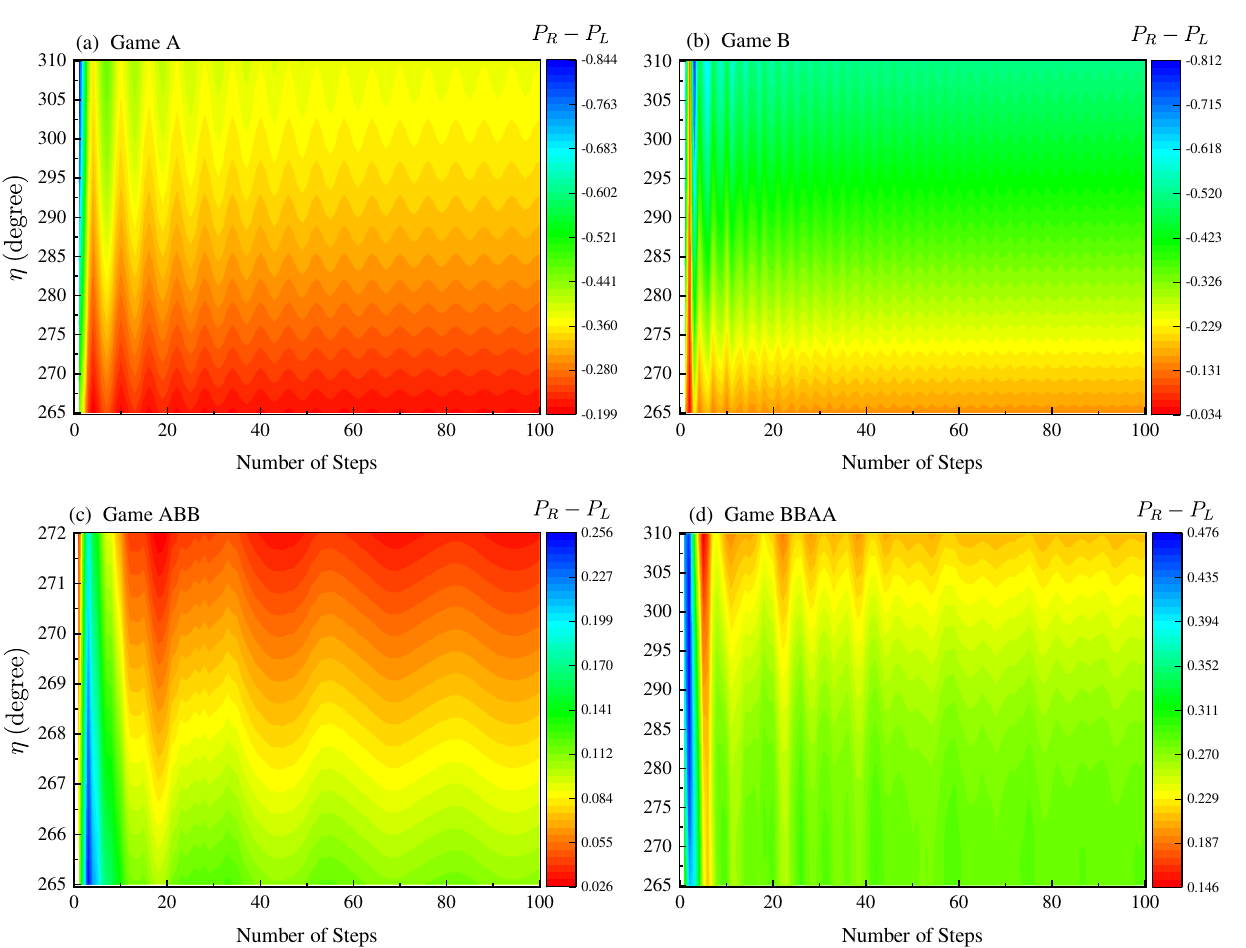}}
    \caption{(color online) Contour dynamics of difference of walker probability distribution $P_{\text{R}}-P_{\text{L}}$ versus number of steps and relative phase of initial state $\eta$ for (a) game AAA with coin operation $C_{\text{A}}(150,30,172)$, (b) game BBB with coin operation $C_{\text{B}}(175,65,165)$, individually. When played alternatively in particular sequences of different periods regarded as different games, e.g., (c) game ABB with coin operation $C_{\text{A}}C_{\text{B}}C_{\text{B}}$, and (d) game BBAA, for a range of $\eta$. These figures demonstrate the occurrence of Parrondo's paradox in 1D discrete-time QWs over infinite number of steps for coins operation $C_{\text{A}}(150,30,172)$ and $C_{\text{B}}(175,65,165)$.}\label{Fig4}
\end{figure*}

\begin{figure*}[ht!]
	{\includegraphics[width=1\linewidth]{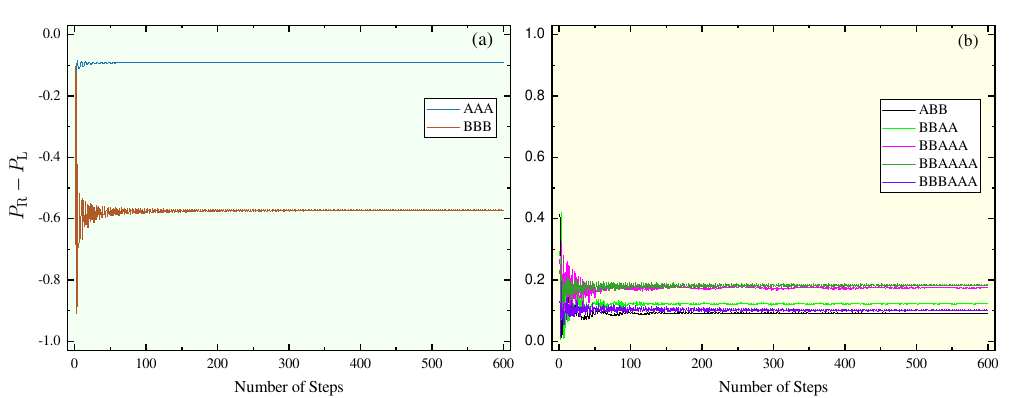}}
    \caption{(color online) Walker difference probability distribution $P_{\text{R}}-P_{\text{L}}$ versus number of steps for (a) game A with coin operation $C_{\text{A}}(155,26,38)$, game B with coin operation $C_{\text{B}}(170,67,118)$, individually. (b) when played alternatively in particular sequences of different periods regarded as different games, e.g., game ABB with coin operation $C_{\text{A}}C_{\text{B}}C_{\text{B}}$, similarly, for BBAA, BBAAA, BBAAAA, BBBAAA, with $\eta=3\pi/2$. These figures display the occurrence of Parrondo's paradox in 1D discrete-time QWs over infinite number of steps for coins operation $C_{\text{A}}(155,26,38)$ and $C_{\text{B}}(170,67,118)$.}\label{Fig5}
\end{figure*}

\begin{figure*}[ht!]
	{\includegraphics[width=1\linewidth]{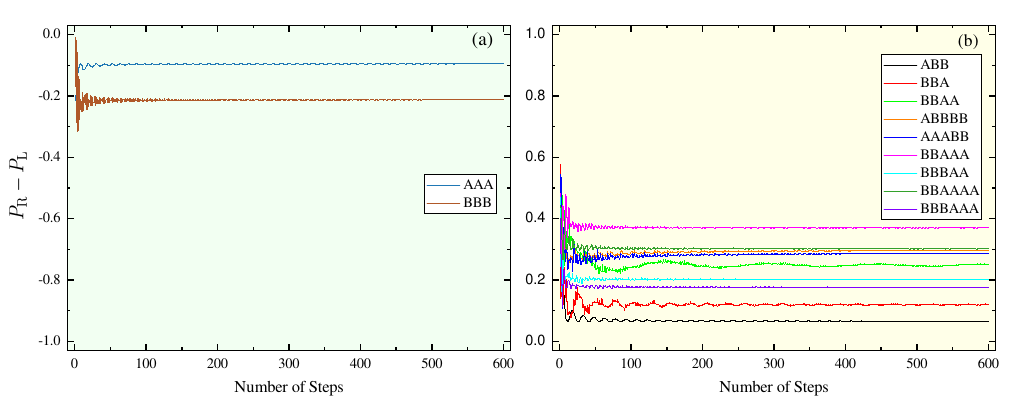}}
    \caption{ (color online) Walker difference probability distribution $P_{\text{R}}-P_{\text{L}}$ versus number of steps for (a) game A with coin operation $C_{\text{A}}(156,16,0)$, game B with coin operation $C_{\text{B}}(0,75,160)$, individually. (b) when played alternatively in particular sequences of different periods regarded as different games, e.g., game ABB with coin operation $C_{\text{A}}C_{\text{B}}C_{\text{B}}$, similarly, for BBA, BBAA, ABBBB, AAABB, BBAAA, BBBAA, BBAAAA, BBBAAA, with $\eta=\pi/2$. These figures display the existence of genuine Parrondo's paradox in 1D discrete-time QWs over infinite number of steps for coins operation $C_{\text{A}}(156,16,0)$ and $C_{\text{B}}(0,75,160)$.} \label{Fig6}
\end{figure*}

\begin{figure*}
	{\includegraphics[width=1\linewidth]{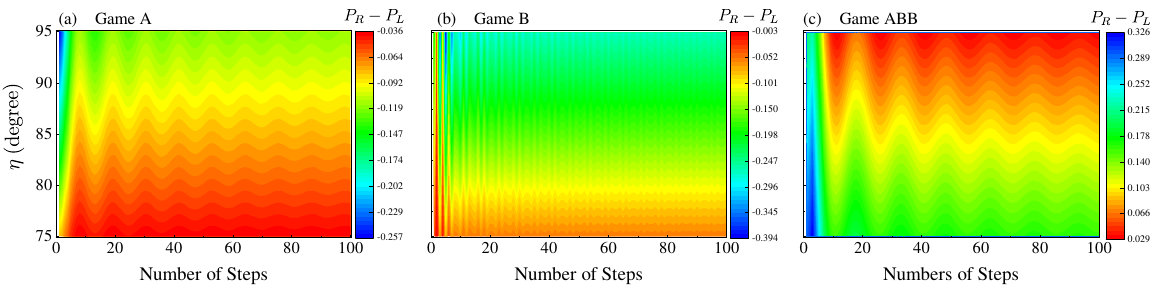}}
    \caption{(color online)  Contour dynamics of difference of walker probability distribution $P_{\text{R}}-P_{\text{L}}$ versus number of steps and relative phase of initial state $\eta$ for (a) game AAA with coin operation $C_{\text{A}}(156,16,0)$, (b) game BBB with coin operation $C_{\text{B}}(0,75,160)$, individually. When played alternatively in particular sequences of different periods regarded as different games, e.g., (c) game ABB with coin operation $C_{\text{A}}C_{\text{B}}C_{\text{B}}$,  for a range of $\eta$. These demonstrate the occurrence of Parrondo's paradox in 1D discrete-time QWs over infinite number of steps for coins operation $C_{\text{A}}(156,16,0)$ and $C_{\text{B}}(0,75,160)$.}\label{Fig7}
\end{figure*}

\begin{figure*}
	{\includegraphics[width=0.98\linewidth]{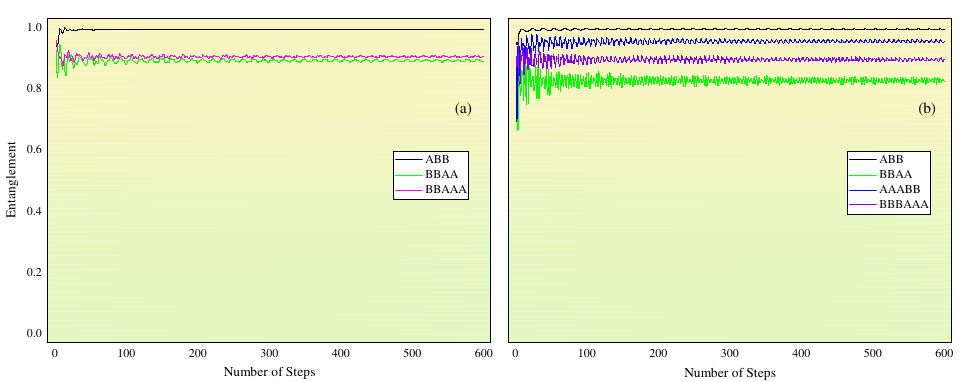}}
    \caption{(color online) Coin-position entanglement versus number of steps for coins operation (a) $C_{\text{A}}(150,30,172)$, $C_{\text{B}}(175,65,165)$, with $\eta=3\pi/2$, and (b) $C_{\text{A}}(156,16,0)$, $C_{\text{B}}(0,75,160)$, with $\eta=\pi/2$, respectively.. Different colors correspond to the particular sequence of different periods, as displayed in the legends. These figures show that coin-position entanglement decreases with increasing number of periods for the corresponding coins operations.} \label{Fig8}
\end{figure*}

\begin{figure*}
	{\includegraphics[width=0.98\linewidth]{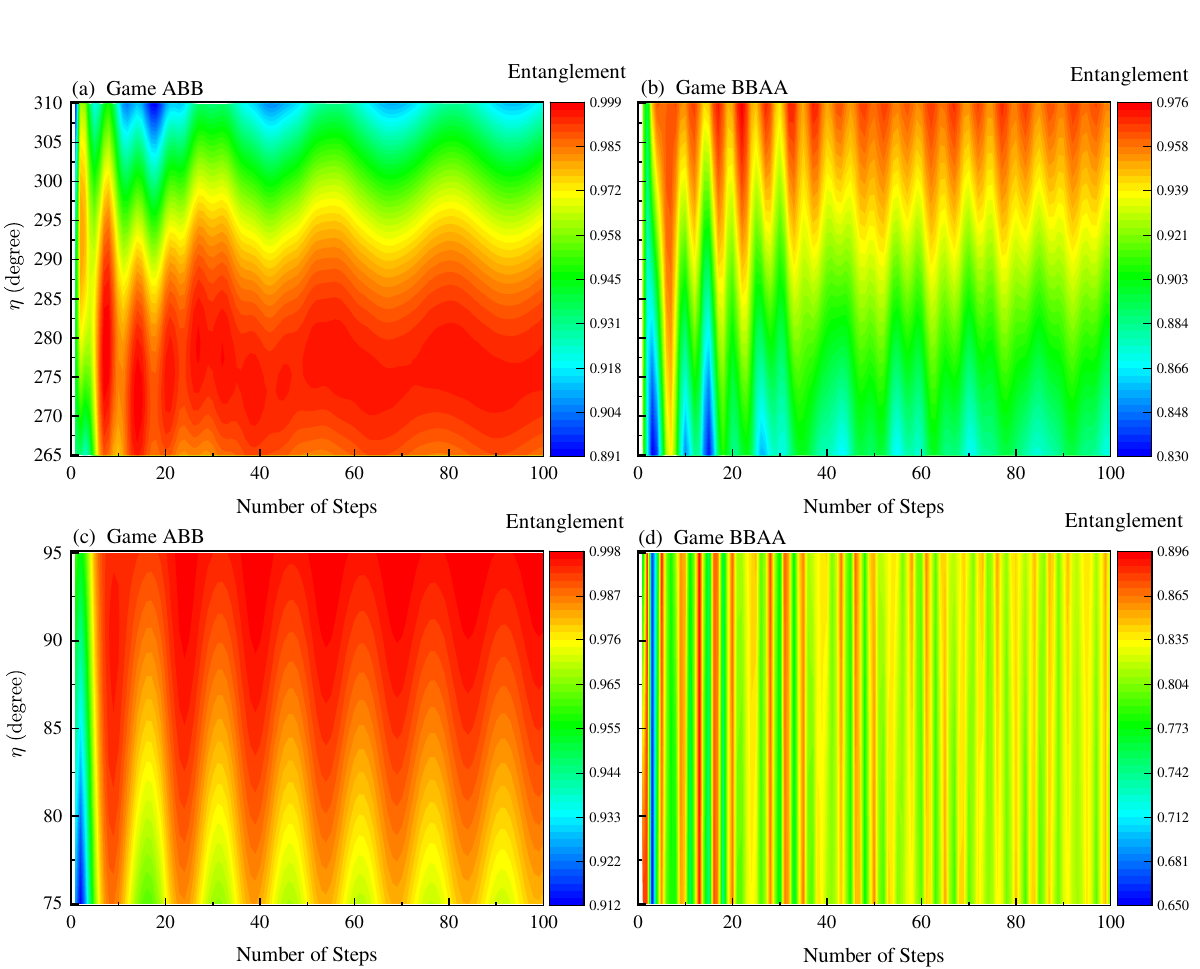}}
    \caption{(color online) Contour dynamics of coin-position entanglement versus number of steps and  relative phase $\eta$ for (a) game ABB (b) game BBAA with coins operator $C_{\text{A}}(150,30,172)$, $C_{\text{B}}(175,65,165)$ and (c) game ABB , (d) game BBAA with coins operator $C_{\text{A}}(156,16,0)$, $C_{\text{B}}(0,75,160)$, respectively. These figures show that Parrondo's sequence of ABB can generate maximal entangler state in our considered scenario.} \label{Fig9}
\end{figure*}

Here, as an example, we discuss three different regions where we play game A with coin rotation operator $C_{\text{A}}$ (orange) and game B with operator $C_{\text{B}}$ (green), as shown in Figs.~\ref{Fig2}$(a)$ and $(b)$, respectively. We show that in all regions when playing games A and B individually at any time, $t$ will lose the game according to the definition. However, when we play these two games alternatively for a particular sequence of different periods, i.e., among which we have only sketched the coin operation pattern for the game ABB (where, we use one-time $C_{\text{A}}$ and two-time $C_{\text{B}}$ for every single step) as shown in the Fig.~\ref{Fig2}$(c)$, can create a winning expectation all the time, known as Parrondo's paradox.   

Firstly, we investigate the dynamics of probability distribution in 1D discrete-time QWs corresponding to games A and B individually. In the game A(B), the quantum state at the step $t$ is $|\Psi(t)\rangle_{\text{A(B)}}=(S{\cdot}C_{\text{A(B)}})^t|\Psi(0)\rangle$. 
The simulation results of the bias of the distribution of the states $\Psi_{\text{A(B)}}(t)$, i.e., $P_{\text{R}}(t)-P_{\text{L}}(t)$ for games A and B, are demonstrated in Fig.~\ref{Fig3}: where Fig.~\ref{Fig3}(a) corresponding to the game A with rotation operator $C_{\text{A}}=C(150,30,172)$ and game B with rotation operator $C_{\text{B}}=C(175,65,165)$. It is clear that both games A and B are losing games and the bias distribution $P_{\text{R}}-P_{\text{L}}$ are negative throughout the number of steps $t$. Whenever we play games A and B alternatively for the particular sequence of different periods, e.g., the game ABB, rotating the coins in the sequence of $C_{\text{A}}C_{\text{B}}C_{\text{B}}$ which has a period of $3$ as shown in Fig.~\ref{Fig2}$(c)$, one can observe some counterintuitive behavior. In this scenario the quantum state at step $t$ is $|\Psi(t)\rangle_{\text{ABB}}=(S{\cdot}C_{\text{B}}{\cdot}S{\cdot}C_{\text{B}}{\cdot}S{\cdot}C_{\text{A}})^t|\Psi(0)\rangle$, the simulation results of the bias distribution of the state at any step $t$ are depicted in Fig.~\ref{Fig3}$(b)$. Similarly, one can also see the same counterintuitive behavior for the sequence of BBAA with coin rotation $C_{\text{B}}C_{\text{B}}C_{\text{A}}C_{\text{A}}$ and BBAAA with coin rotation $C_{\text{B}}C_{\text{B}}C_{\text{A}}C_{\text{A}}C_{\text{A}}$ for the period of $4$ and $5$, with $\eta=3\pi/2$, respectively. The results are shown in the Fig.~\ref{Fig3}$(b)$, where the different colors correspond to different games, as displayed in the figures.

Further, to understand the dependency of the relative phase of initial state of coins $\eta$ we have figured out the range of $\eta$ where Parrondo's paradox exist in our considered scheme. In case of rotation operators $C_{\text{A}}=C(150,30,172)$ and $C_{\text{B}}=C(175,65,165)$, one can see that for $\eta= 265^{\circ}-310^{\circ}$ the games A and B, are losing games according to the above defined strategy by displaying negative $P_{\text{R}}-P_{\text{L}}$. Similarly, games ABB and BBAA are the winning games by exhibiting the positive $P_{\text{R}}-P_{\text{L}}$, dynamics as shown in Figs.~\ref{Fig5}$(a)-(d)$, respectively. We limited our simulation up to 100 number of steps in order to show the dynamics clearly but this dynamics will be similar for an infinite number of steps and the range of the $\eta$ will change with varying the rotation operators of coins.

Likewise, we also demonstrate the paradoxical behavior for the other coin parameters regime where game A play with coin rotation operator $C_{\text{A}}=C(155,26,38)$ and game B with $C_{\text{B}}=C(170,67,118)$ for which $P_{\text{R}}-P_{\text{L}}$ dynamics is throughout negative, representing the losing games, when playing these games individually for any time, $t$, as shown in Fig.~\ref{Fig4}$(a)$. Whereas playing with both coins simultaneously for the different sequences like, ABB, BBAA, BBAAA, BBAAAA, BBBAAA, with $\eta=3\pi/2$, one can see the dynamics of $P_{\text{R}}-P_{\text{L}}$, which is positive throughout the time evolution, showing the paradoxical scenario, as depicted in Fig.~\ref{Fig4}$(b)$.

Moreover, we discuss the scenario of the different initial state and biasing of coin parameters. In the first two cases, we have biased the coin A and B on both sides of the origin, meaning that both $\alpha$ and $\gamma$, are non-zero, representing the simultaneous biasing situation. Here, we consider the scenario of $\gamma_{\text{A}}=0$ ($\alpha_{\text{B}}=0$), for the coin A (B), respectively. Now using the coin parameters $C_{\text{A}}=C(156,16,0)$, and $C_{\text{B}}=C(0,75,160)$, we show that games A and B are losing games when played individually for any number of steps $t$, as displayed in Fig.~\ref{Fig6}$(a)$. But whenever we play alternatively with both of these coins, one can observe the paradoxical behavior for some of the particular sequence in almost every period. Limited our calculation up-to period six, we show that the sequence ABB, BBA, BBAA, ABBBB, AAABB, BBAAA, BBBAA, BBAAAA, BBBAAA, with $\eta=\pi/2$, lead to a winning expectation, as shown by different colors in Fig.~\ref{Fig6}$(b)$. Moreover, it is clear that generally, increasing the periods can enhance the outcome of the winning games, sometime even or odd period will generate maximum winning outcomes, depending strongly on the biasing parameters.  According to the definition of losing and winning strategies of the Parrondo's game in the QW, we have demonstrated that combining two losing games (A and B) in all of our three scenarios, can produce a winning game called Parrondo's paradox. Moreover, the set of coin rotation parameters for which we observe the Parrondo's paradox are not special in any way and one can find many such sets of parameters where Parrondo’s effect can be observed.

In addition, to discuss the dependency of the relative phase of initial state of coins $\eta$ we have figured out the range of $\eta$ where the Parrondo's paradox exists in our considered scheme. In case of rotation operators $C_{\text{A}}=C(156,16,0)$ and $C_{\text{B}}=C(0,75,160)$, one can see that for $\eta= 75^{\circ}-95^{\circ}$ games A and B, are losing games by displaying negative $P_{\text{R}}-P_{\text{L}}$ dynamics and game ABB is the winning game by exhibiting the positive $P_{\text{R}}-P_{\text{L}}$, dynamics as shown in Figs.~\ref{Fig7}$(a)-(c)$, respectively. We limited our simulation up to 100 number of steps in order to show the dynamics clearly but this behavior will be hold for an infinite number of steps and the range of $\eta$ will change with varying the rotation operators of coins.

Parrondo's games can be viewed as a special case of disordered QWs, which can enhance the entanglement generation between the coin and position\,\cite{PhysRevLett.111.180503,wang2018dynamic}. In order to calculate the coin-position entanglement numerically we use the von Neumann entropy $S_E(\rho(t)) = -\text{Tr}[\rho_C(t)\log_2\rho_C(t)]$, where $\rho_C(t) = \text{Tr}[\rho(t)]$ is the reduced density matrix of the coin and $\rho(t) = |\Psi(t)\rangle\langle\Psi(t)|$ represents the global density matrix of the walker by assuming that the system in a pure state. Here, $S_E = 0$ and $1$ correspond to the separable and maximally entangled states, respectively. Comparing the entanglement dynamics of Fig.~\ref{Fig8}$(a)$ with probability distribution of Fig.~\ref{Fig3}$(b)$, one can see that increasing period for the corresponding coin operations can enhance the asymmetry in the probability distribution of the walker toward the right of the origin, leading to a maximum winning outcome. The sequence which possesses maximum winning outcomes, e.g., BBAA in Fig.~\ref{Fig3}$(b)$, may generate minimal entanglement than that which possesses minimal winning outcome, like ABB, generate maximal entanglement for the same coin operators. Similarly, Fig.~\ref{Fig8}$(b)$ demonstrate the entanglement dynamics of different periods of Fig.~\ref{Fig6}$(b)$, where one can see that the sequence ABB possesses maximum entanglement than all other sequences. 

Further, to explore the range of the relative phase of initial state $\eta$ where Parrondo's sequence may generate maximal entangler state is different for distinct coins operator. One can observe that in the case of $C_{\text{A}}(150,30,172)$, and $C_{\text{B}}(175,65,165)$ we have achieved maximal entangler state of the value $0.999$ for the sequence of ABB, and $0.976$ in the case of BBAA, which can be seen in the entanglement dynamics in Figs.~\ref{Fig9}$(a)-(b)$, respectively. Similary, in case of  $C_{\text{A}}(156,16,0)$ and $C_{\text{B}}(0,75,160)$, one can obtain maximal entanglement of $0.998$ and $0.896$ for the sequence of ABB and BBAA, as shown in the entanglement dynamics in Figs.~\ref{Fig9}$(c)-(d)$, respectively. From all these discussions, it becomes clear that Parrondo game is very sensitive to the biasing, relative phase and its initial state of coins, but in case of QWs, the interference phenomena also play a key role in this counterintuitive behavior.

In summary, we have demonstrated the scenario of  the \emph{quantum} Parrondo effect in 1D discrete-time QWs using different coin operators regraded as a different games. By measuring the mean position of the walker in its final step, we clearly show that two losing strategies can win by playing the two games in a particular periodic sequence, known as Parrondo's paradox. Here, as an example we have demonstrated three different regions of coin operators where this paradox exists and these sets of parameters are not special, one can find many such sets of parameters for the which the genuine Parrondo's effect may exist. We have displayed regimes of the relative phases of the initial state of coins where Parrondo games exist. We have also discussed the coin-position entanglement generation in QWs for the different choices of relative phase of the initial state of coins and found that the Parrondo's sequence of the period three (ABB) can generate maximal entanglement than all other sequences but the exact relations of Parrondo's games with the entanglement and coherence require further study. Furthermore, we also found that the quantum interference plays an important role in such a quantum counterpart of Parrondo effect. The Parrondo's game supplies a new insight for the alternative QWS and we hope it will be helpful to develop new quantum algorithms.

\section*{Acknowledgment}
M. Jan acknowledges the Grant No. ZC304022918 to support
his Postdoctoral Fellowship at Zhejiang Normal University. G.X. acknowledges support from the NSFC under Grants No. 11835011 and No. 11774316.

\providecommand{\noopsort}[1]{}\providecommand{\singleletter}[1]{#1}%
%



\begin{thebibliography}{59}%
\makeatletter
\providecommand \@ifxundefined [1]{%
 \@ifx{#1\undefined}
}%
\providecommand \@ifnum [1]{%
 \ifnum #1\expandafter \@firstoftwo
 \else \expandafter \@secondoftwo
 \fi
}%
\providecommand \@ifx [1]{%
 \ifx #1\expandafter \@firstoftwo
 \else \expandafter \@secondoftwo
 \fi
}%
\providecommand \natexlab [1]{#1}%
\providecommand \enquote  [1]{``#1''}%
\providecommand \bibnamefont  [1]{#1}%
\providecommand \bibfnamefont [1]{#1}%
\providecommand \citenamefont [1]{#1}%
\providecommand \href@noop [0]{\@secondoftwo}%
\providecommand \href [0]{\begingroup \@sanitize@url \@href}%
\providecommand \@href[1]{\@@startlink{#1}\@@href}%
\providecommand \@@href[1]{\endgroup#1\@@endlink}%
\providecommand \@sanitize@url [0]{\catcode `\\12\catcode `\$12\catcode
  `\&12\catcode `\#12\catcode `\^12\catcode `\_12\catcode `\%12\relax}%
\providecommand \@@startlink[1]{}%
\providecommand \@@endlink[0]{}%
\providecommand \url  [0]{\begingroup\@sanitize@url \@url }%
\providecommand \@url [1]{\endgroup\@href {#1}{\urlprefix }}%
\providecommand \urlprefix  [0]{URL }%
\providecommand \Eprint [0]{\href }%
\providecommand \doibase [0]{http://dx.doi.org/}%
\providecommand \selectlanguage [0]{\@gobble}%
\providecommand \bibinfo  [0]{\@secondoftwo}%
\providecommand \bibfield  [0]{\@secondoftwo}%
\providecommand \translation [1]{[#1]}%
\providecommand \BibitemOpen [0]{}%
\providecommand \bibitemStop [0]{}%
\providecommand \bibitemNoStop [0]{.\EOS\space}%
\providecommand \EOS [0]{\spacefactor3000\relax}%
\providecommand \BibitemShut  [1]{\csname bibitem#1\endcsname}%
\let\auto@bib@innerbib\@empty
\bibitem [{\citenamefont {Parrondo}(1996)}]{parrondo1996eec}%
  \BibitemOpen
  \bibfield  {author} {\bibinfo {author} {\bibfnamefont {J.~M.~R.}\
  \bibnamefont {Parrondo}},\ }\href@noop {} {\enquote {\bibinfo {title} {How to
  cheat a bad mathematician},}\ } (\bibinfo {year} {1996}),\ \bibinfo {note}
  {in EEC HC\&M Network on Complexity and Chaos (\#ERBCHRX-CT940546),
  Unpublished (1996)}\BibitemShut {NoStop}%
\bibitem [{\citenamefont {Parrondo}\ \emph {et~al.}(2000)\citenamefont
  {Parrondo}, \citenamefont {Harmer},\ and\ \citenamefont
  {Abbott}}]{parrondo2000new}%
  \BibitemOpen
  \bibfield  {author} {\bibinfo {author} {\bibfnamefont {J.~M.~R.}\
  \bibnamefont {Parrondo}}, \bibinfo {author} {\bibfnamefont {G.~P.}\
  \bibnamefont {Harmer}}, \ and\ \bibinfo {author} {\bibfnamefont
  {D.}~\bibnamefont {Abbott}},\ }\href {\doibase 10.1103/PhysRevLett.85.5226}
  {\bibfield  {journal} {\bibinfo  {journal} {Phys. Rev. Lett.}\ }\textbf
  {\bibinfo {volume} {85}},\ \bibinfo {pages} {5226} (\bibinfo {year}
  {2000})}\BibitemShut {NoStop}%
\bibitem [{\citenamefont {Harmer}\ and\ \citenamefont
  {Abbott}(1999{\natexlab{a}})}]{harmer1999game}%
  \BibitemOpen
  \bibfield  {author} {\bibinfo {author} {\bibfnamefont {G.~P.}\ \bibnamefont
  {Harmer}}\ and\ \bibinfo {author} {\bibfnamefont {D.}~\bibnamefont
  {Abbott}},\ }\href {\doibase 10.1038/47220} {\bibfield  {journal} {\bibinfo
  {journal} {Nature}\ }\textbf {\bibinfo {volume} {402}},\ \bibinfo {pages}
  {864} (\bibinfo {year} {1999}{\natexlab{a}})}\BibitemShut {NoStop}%
\bibitem [{\citenamefont {Harmer}\ and\ \citenamefont
  {Abbott}(1999{\natexlab{b}})}]{harmer1999parrondo}%
  \BibitemOpen
  \bibfield  {author} {\bibinfo {author} {\bibfnamefont {G.~P.}\ \bibnamefont
  {Harmer}}\ and\ \bibinfo {author} {\bibfnamefont {D.}~\bibnamefont
  {Abbott}},\ }\href {\doibase 10.1214/ss/1009212247} {\bibfield  {journal}
  {\bibinfo  {journal} {Stat. Sci.}\ }\textbf {\bibinfo {volume} {14}},\
  \bibinfo {pages} {206} (\bibinfo {year} {1999}{\natexlab{b}})}\BibitemShut
  {NoStop}%
\bibitem [{\citenamefont {Jian-Jun}\ and\ \citenamefont
  {Qi-Wen}(2014)}]{jian2014beyond}%
  \BibitemOpen
  \bibfield  {author} {\bibinfo {author} {\bibfnamefont {S.}~\bibnamefont
  {Jian-Jun}}\ and\ \bibinfo {author} {\bibfnamefont {W.}~\bibnamefont
  {Qi-Wen}},\ }\href {\doibase 10.1038/srep04244} {\bibfield  {journal}
  {\bibinfo  {journal} {Sci. Rep.}\ }\textbf {\bibinfo {volume} {4}},\ \bibinfo
  {pages} {4244} (\bibinfo {year} {2014})}\BibitemShut {NoStop}%
\bibitem [{\citenamefont {Parrondo}\ and\ \citenamefont
  {Dinís}(2004)}]{parrondo2004brownian}%
  \BibitemOpen
  \bibfield  {author} {\bibinfo {author} {\bibfnamefont {J.~M.~R.}\
  \bibnamefont {Parrondo}}\ and\ \bibinfo {author} {\bibfnamefont
  {L.}~\bibnamefont {Dinís}},\ }\href {\doibase 10.1080/00107510310001644836}
  {\bibfield  {journal} {\bibinfo  {journal} {Contemp. Phys.}\ }\textbf
  {\bibinfo {volume} {45}},\ \bibinfo {pages} {147} (\bibinfo {year}
  {2004})}\BibitemShut {NoStop}%
\bibitem [{\citenamefont {Rosato}\ \emph {et~al.}(1987)\citenamefont {Rosato},
  \citenamefont {Strandburg}, \citenamefont {Prinz},\ and\ \citenamefont
  {Swendsen}}]{PhysRevLett.58.1038}%
  \BibitemOpen
  \bibfield  {author} {\bibinfo {author} {\bibfnamefont {A.}~\bibnamefont
  {Rosato}}, \bibinfo {author} {\bibfnamefont {K.~J.}\ \bibnamefont
  {Strandburg}}, \bibinfo {author} {\bibfnamefont {F.}~\bibnamefont {Prinz}}, \
  and\ \bibinfo {author} {\bibfnamefont {R.~H.}\ \bibnamefont {Swendsen}},\
  }\href {\doibase 10.1103/PhysRevLett.58.1038} {\bibfield  {journal} {\bibinfo
   {journal} {Phys. Rev. Lett.}\ }\textbf {\bibinfo {volume} {58}},\ \bibinfo
  {pages} {1038} (\bibinfo {year} {1987})}\BibitemShut {NoStop}%
\bibitem [{\citenamefont {Westerhoff}\ \emph {et~al.}(1986)\citenamefont
  {Westerhoff}, \citenamefont {Tsong}, \citenamefont {Chock}, \citenamefont
  {Chen},\ and\ \citenamefont {Astumian}}]{westerhoff1986enzymes}%
  \BibitemOpen
  \bibfield  {author} {\bibinfo {author} {\bibfnamefont {H.~V.}\ \bibnamefont
  {Westerhoff}}, \bibinfo {author} {\bibfnamefont {T.~Y.}\ \bibnamefont
  {Tsong}}, \bibinfo {author} {\bibfnamefont {P.~B.}\ \bibnamefont {Chock}},
  \bibinfo {author} {\bibfnamefont {Y.-D.}\ \bibnamefont {Chen}}, \ and\
  \bibinfo {author} {\bibfnamefont {R.}~\bibnamefont {Astumian}},\ }\href@noop
  {} {\bibfield  {journal} {\bibinfo  {journal} {Proc. Natl. Acad. Sci.}\
  }\textbf {\bibinfo {volume} {83}},\ \bibinfo {pages} {4734} (\bibinfo {year}
  {1986})}\BibitemShut {NoStop}%
\bibitem [{\citenamefont {Key}(1987)}]{key1987computable}%
  \BibitemOpen
  \bibfield  {author} {\bibinfo {author} {\bibfnamefont {E.}~\bibnamefont
  {Key}},\ }\href {\doibase 10.1007/BF00320084} {\bibfield  {journal} {\bibinfo
   {journal} {Probab. Theory Relat. Fields}\ }\textbf {\bibinfo {volume}
  {75}},\ \bibinfo {pages} {97} (\bibinfo {year} {1987})}\BibitemShut {NoStop}%
\bibitem [{\citenamefont {Maslov}\ and\ \citenamefont
  {Zhang}(1998)}]{maslov1998optimal}%
  \BibitemOpen
  \bibfield  {author} {\bibinfo {author} {\bibfnamefont {S.}~\bibnamefont
  {Maslov}}\ and\ \bibinfo {author} {\bibfnamefont {Y.-C.}\ \bibnamefont
  {Zhang}},\ }\href {\doibase 10.1142/S0219024998000217} {\bibfield  {journal}
  {\bibinfo  {journal} {Int. J. Theor. Appl. Finance}\ }\textbf {\bibinfo
  {volume} {1}},\ \bibinfo {pages} {377} (\bibinfo {year} {1998})}\BibitemShut
  {NoStop}%
\bibitem [{\citenamefont {Flitney}\ \emph {et~al.}(2002)\citenamefont
  {Flitney}, \citenamefont {Ng},\ and\ \citenamefont
  {Abbott}}]{flitney2002quantum}%
  \BibitemOpen
  \bibfield  {author} {\bibinfo {author} {\bibfnamefont {A.~P.}\ \bibnamefont
  {Flitney}}, \bibinfo {author} {\bibfnamefont {J.}~\bibnamefont {Ng}}, \ and\
  \bibinfo {author} {\bibfnamefont {D.}~\bibnamefont {Abbott}},\ }\href
  {\doibase 10.1016/S0378-4371(02)01084-1} {\bibfield  {journal} {\bibinfo
  {journal} {Physica A}\ }\textbf {\bibinfo {volume} {314}},\ \bibinfo {pages}
  {35} (\bibinfo {year} {2002})}\BibitemShut {NoStop}%
\bibitem [{\citenamefont {Lee}\ \emph {et~al.}(2002)\citenamefont {Lee},
  \citenamefont {Johnson}, \citenamefont {Rodriguez},\ and\ \citenamefont
  {Quiroga}}]{lee2002quantum}%
  \BibitemOpen
  \bibfield  {author} {\bibinfo {author} {\bibfnamefont {C.~F.}\ \bibnamefont
  {Lee}}, \bibinfo {author} {\bibfnamefont {N.~F.}\ \bibnamefont {Johnson}},
  \bibinfo {author} {\bibfnamefont {F.}~\bibnamefont {Rodriguez}}, \ and\
  \bibinfo {author} {\bibfnamefont {L.}~\bibnamefont {Quiroga}},\ }\href
  {\doibase 10.1142/S0219477502000920} {\bibfield  {journal} {\bibinfo
  {journal} {Fluctuation Noise Lett.}\ }\textbf {\bibinfo {volume} {2}},\
  \bibinfo {pages} {293} (\bibinfo {year} {2002})}\BibitemShut {NoStop}%
\bibitem [{\citenamefont {Buceta}\ \emph {et~al.}(2001)\citenamefont {Buceta},
  \citenamefont {Lindenberg},\ and\ \citenamefont
  {Parrondo}}]{PhysRevLett.88.024103}%
  \BibitemOpen
  \bibfield  {author} {\bibinfo {author} {\bibfnamefont {J.}~\bibnamefont
  {Buceta}}, \bibinfo {author} {\bibfnamefont {K.}~\bibnamefont {Lindenberg}},
  \ and\ \bibinfo {author} {\bibfnamefont {J.~M.~R.}\ \bibnamefont
  {Parrondo}},\ }\href {\doibase 10.1103/PhysRevLett.88.024103} {\bibfield
  {journal} {\bibinfo  {journal} {Phys. Rev. Lett.}\ }\textbf {\bibinfo
  {volume} {88}},\ \bibinfo {pages} {024103} (\bibinfo {year}
  {2001})}\BibitemShut {NoStop}%
\bibitem [{\citenamefont {Khan}\ \emph {et~al.}(2010)\citenamefont {Khan},
  \citenamefont {Ramzan},\ and\ \citenamefont {Khan}}]{khan2010quantum}%
  \BibitemOpen
  \bibfield  {author} {\bibinfo {author} {\bibfnamefont {S.}~\bibnamefont
  {Khan}}, \bibinfo {author} {\bibfnamefont {M.}~\bibnamefont {Ramzan}}, \ and\
  \bibinfo {author} {\bibfnamefont {M.~K.}\ \bibnamefont {Khan}},\ }\href
  {\doibase 10.1007/s10773-009-0175-y} {\bibfield  {journal} {\bibinfo
  {journal} {Int. J. Theor. Phys.}\ }\textbf {\bibinfo {volume} {49}},\
  \bibinfo {pages} {31} (\bibinfo {year} {2010})}\BibitemShut {NoStop}%
\bibitem [{\citenamefont {Ethier}\ and\ \citenamefont
  {Lee}(2012)}]{ethier2012parrondo}%
  \BibitemOpen
  \bibfield  {author} {\bibinfo {author} {\bibfnamefont {S.}~\bibnamefont
  {Ethier}}\ and\ \bibinfo {author} {\bibfnamefont {J.}~\bibnamefont {Lee}},\
  }\href@noop {} {\bibfield  {journal} {\bibinfo  {journal} {arXiv:1203.0818
  [math.PR]}\ } (\bibinfo {year} {2012})}\BibitemShut {NoStop}%
\bibitem [{\citenamefont {Pawela}\ and\ \citenamefont
  {S{\l}adkowski}(2013)}]{pawela2013cooperative}%
  \BibitemOpen
  \bibfield  {author} {\bibinfo {author} {\bibfnamefont {{\L}.}~\bibnamefont
  {Pawela}}\ and\ \bibinfo {author} {\bibfnamefont {J.}~\bibnamefont
  {S{\l}adkowski}},\ }\href {\doibase 10.1016/j.physd.2013.04.010} {\bibfield
  {journal} {\bibinfo  {journal} {Physica D}\ }\textbf {\bibinfo {volume}
  {256}},\ \bibinfo {pages} {51} (\bibinfo {year} {2013})}\BibitemShut
  {NoStop}%
\bibitem [{\citenamefont {Pejic}(2015)}]{pejic2015quantum}%
  \BibitemOpen
  \bibfield  {author} {\bibinfo {author} {\bibfnamefont {M.}~\bibnamefont
  {Pejic}},\ }\href@noop {} {\bibfield  {journal} {\bibinfo  {journal}
  {arXiv:1503.08868 [math-ph]}\ } (\bibinfo {year} {2015})}\BibitemShut
  {NoStop}%
\bibitem [{\citenamefont {Gr{\"u}nbaum}\ and\ \citenamefont
  {Pejic}(2016)}]{grunbaum2016maximal}%
  \BibitemOpen
  \bibfield  {author} {\bibinfo {author} {\bibfnamefont {F.~A.}\ \bibnamefont
  {Gr{\"u}nbaum}}\ and\ \bibinfo {author} {\bibfnamefont {M.}~\bibnamefont
  {Pejic}},\ }\href {\doibase 10.1007/s11005-015-0812-8} {\bibfield  {journal}
  {\bibinfo  {journal} {Lett. Math. Phys.}\ }\textbf {\bibinfo {volume}
  {106}},\ \bibinfo {pages} {251} (\bibinfo {year} {2016})}\BibitemShut
  {NoStop}%
\bibitem [{\citenamefont {Masuda}\ and\ \citenamefont
  {Konno}(2004)}]{masuda2004subcritical}%
  \BibitemOpen
  \bibfield  {author} {\bibinfo {author} {\bibfnamefont {N.}~\bibnamefont
  {Masuda}}\ and\ \bibinfo {author} {\bibfnamefont {N.}~\bibnamefont {Konno}},\
  }\href {\doibase 10.1140/epjb/e2004-00279-5} {\bibfield  {journal} {\bibinfo
  {journal} {Eur. Phys. J. B}\ }\textbf {\bibinfo {volume} {40}},\ \bibinfo
  {pages} {313} (\bibinfo {year} {2004})}\BibitemShut {NoStop}%
\bibitem [{\citenamefont {Wolf}\ \emph {et~al.}(2005)\citenamefont {Wolf},
  \citenamefont {Vazirani},\ and\ \citenamefont {Arkin}}]{wolf2005diversity}%
  \BibitemOpen
  \bibfield  {author} {\bibinfo {author} {\bibfnamefont {D.~M.}\ \bibnamefont
  {Wolf}}, \bibinfo {author} {\bibfnamefont {V.~V.}\ \bibnamefont {Vazirani}},
  \ and\ \bibinfo {author} {\bibfnamefont {A.~P.}\ \bibnamefont {Arkin}},\
  }\href {\doibase 10.1016/j.jtbi.2004.11.020} {\bibfield  {journal} {\bibinfo
  {journal} {J. Theor. Biol.}\ }\textbf {\bibinfo {volume} {234}},\ \bibinfo
  {pages} {227} (\bibinfo {year} {2005})}\BibitemShut {NoStop}%
\bibitem [{\citenamefont {Reed}(2007)}]{reed2007two}%
  \BibitemOpen
  \bibfield  {author} {\bibinfo {author} {\bibfnamefont {F.~A.}\ \bibnamefont
  {Reed}},\ }\href {\doibase 10.1534/genetics.106.069997} {\bibfield  {journal}
  {\bibinfo  {journal} {Genetics}\ }\textbf {\bibinfo {volume} {176}},\
  \bibinfo {pages} {1923} (\bibinfo {year} {2007})}\BibitemShut {NoStop}%
\bibitem [{\citenamefont {Spurgin}\ and\ \citenamefont
  {Tamarkin}(2005)}]{spurgin2005switching}%
  \BibitemOpen
  \bibfield  {author} {\bibinfo {author} {\bibfnamefont {R.}~\bibnamefont
  {Spurgin}}\ and\ \bibinfo {author} {\bibfnamefont {M.}~\bibnamefont
  {Tamarkin}},\ }\href {\doibase 10.1207/s15427579jpfm0601_3} {\bibfield
  {journal} {\bibinfo  {journal} {J. Behav. Finance}\ }\textbf {\bibinfo
  {volume} {6}},\ \bibinfo {pages} {15} (\bibinfo {year} {2005})}\BibitemShut
  {NoStop}%
\bibitem [{\citenamefont {Amengual}\ \emph {et~al.}(2004)\citenamefont
  {Amengual}, \citenamefont {Allison}, \citenamefont {Toral},\ and\
  \citenamefont {Abbott}}]{Amengual2004}%
  \BibitemOpen
  \bibfield  {author} {\bibinfo {author} {\bibfnamefont {P.}~\bibnamefont
  {Amengual}}, \bibinfo {author} {\bibfnamefont {A.}~\bibnamefont {Allison}},
  \bibinfo {author} {\bibfnamefont {R.}~\bibnamefont {Toral}}, \ and\ \bibinfo
  {author} {\bibfnamefont {D.}~\bibnamefont {Abbott}},\ }\href {\doibase
  10.1098/rspa.2004.1283} {\bibfield  {journal} {\bibinfo  {journal} {Proc.
  Royal Society London A}\ }\textbf {\bibinfo {volume} {460}},\ \bibinfo
  {pages} {2269} (\bibinfo {year} {2004})}\BibitemShut {NoStop}%
\bibitem [{\citenamefont {Cannata}\ \emph {et~al.}(1999)\citenamefont
  {Cannata}, \citenamefont {Ioffe}, \citenamefont {Junker},\ and\ \citenamefont
  {Nishnianidze}}]{Cannata_1999}%
  \BibitemOpen
  \bibfield  {author} {\bibinfo {author} {\bibfnamefont {F.}~\bibnamefont
  {Cannata}}, \bibinfo {author} {\bibfnamefont {M.}~\bibnamefont {Ioffe}},
  \bibinfo {author} {\bibfnamefont {G.}~\bibnamefont {Junker}}, \ and\ \bibinfo
  {author} {\bibfnamefont {D.}~\bibnamefont {Nishnianidze}},\ }\href {\doibase
  10.1088/0305-4470/32/19/309} {\bibfield  {journal} {\bibinfo  {journal} {J.
  Phys. A: Math. Gen.}\ }\textbf {\bibinfo {volume} {32}},\ \bibinfo {pages}
  {3583} (\bibinfo {year} {1999})}\BibitemShut {NoStop}%
\bibitem [{\citenamefont {Aharonov}\ \emph {et~al.}(1993)\citenamefont
  {Aharonov}, \citenamefont {Davidovich},\ and\ \citenamefont
  {Zagury}}]{PhysRevA.48.1687}%
  \BibitemOpen
  \bibfield  {author} {\bibinfo {author} {\bibfnamefont {Y.}~\bibnamefont
  {Aharonov}}, \bibinfo {author} {\bibfnamefont {L.}~\bibnamefont
  {Davidovich}}, \ and\ \bibinfo {author} {\bibfnamefont {N.}~\bibnamefont
  {Zagury}},\ }\href {\doibase 10.1103/PhysRevA.48.1687} {\bibfield  {journal}
  {\bibinfo  {journal} {Phys. Rev. A}\ }\textbf {\bibinfo {volume} {48}},\
  \bibinfo {pages} {1687} (\bibinfo {year} {1993})}\BibitemShut {NoStop}%
\bibitem [{\citenamefont {Venegas-Andraca}(2012)}]{venegas2012quantum}%
  \BibitemOpen
  \bibfield  {author} {\bibinfo {author} {\bibfnamefont {S.~E.}\ \bibnamefont
  {Venegas-Andraca}},\ }\href {\doibase 10.1007/s11128-012-0432-5} {\bibfield
  {journal} {\bibinfo  {journal} {Quantum Inf. Process}\ }\textbf {\bibinfo
  {volume} {11}},\ \bibinfo {pages} {1015} (\bibinfo {year}
  {2012})}\BibitemShut {NoStop}%
\bibitem [{\citenamefont {Mackay}\ \emph {et~al.}(2002)\citenamefont {Mackay},
  \citenamefont {Bartlett}, \citenamefont {Stephenson},\ and\ \citenamefont
  {Sanders}}]{mackay2002quantum}%
  \BibitemOpen
  \bibfield  {author} {\bibinfo {author} {\bibfnamefont {T.~D.}\ \bibnamefont
  {Mackay}}, \bibinfo {author} {\bibfnamefont {S.~D.}\ \bibnamefont
  {Bartlett}}, \bibinfo {author} {\bibfnamefont {L.~T.}\ \bibnamefont
  {Stephenson}}, \ and\ \bibinfo {author} {\bibfnamefont {B.~C.}\ \bibnamefont
  {Sanders}},\ }\href {\doibase 10.1088/0305-4470/35/12/304} {\bibfield
  {journal} {\bibinfo  {journal} {J. Phys. A: Math. Gen.}\ }\textbf {\bibinfo
  {volume} {35}},\ \bibinfo {pages} {2745} (\bibinfo {year}
  {2002})}\BibitemShut {NoStop}%
\bibitem [{\citenamefont {Santha}(2008)}]{Santha2008}%
  \BibitemOpen
  \bibfield  {author} {\bibinfo {author} {\bibfnamefont {M.}~\bibnamefont
  {Santha}},\ }in\ \href@noop {} {\emph {\bibinfo {booktitle} {Theory and
  Applications of Models of Computation}}},\ \bibinfo {editor} {edited by\
  \bibinfo {editor} {\bibfnamefont {M.}~\bibnamefont {Agrawal}}, \bibinfo
  {editor} {\bibfnamefont {D.}~\bibnamefont {Du}}, \bibinfo {editor}
  {\bibfnamefont {Z.}~\bibnamefont {Duan}}, \ and\ \bibinfo {editor}
  {\bibfnamefont {A.}~\bibnamefont {Li}}}\ (\bibinfo  {publisher} {Springer
  Berlin Heidelberg},\ \bibinfo {year} {2008})\ p.~\bibinfo {pages}
  {31}\BibitemShut {NoStop}%
\bibitem [{\citenamefont {Portugal}(2013)}]{Portugal2013}%
  \BibitemOpen
  \bibfield  {author} {\bibinfo {author} {\bibfnamefont {R.}~\bibnamefont
  {Portugal}},\ }\href {\doibase 10.1007/978-1-4614-6336-8} {\emph {\bibinfo
  {title} {Quantum Walks and Search Algorithms}}},\ Quantum Science and
  Technology\ (\bibinfo  {publisher} {Springer-Verlag New York},\ \bibinfo
  {year} {2013})\BibitemShut {NoStop}%
\bibitem [{\citenamefont {Childs}\ and\ \citenamefont {Ge}(2014)}]{Childs2014}%
  \BibitemOpen
  \bibfield  {author} {\bibinfo {author} {\bibfnamefont {A.~M.}\ \bibnamefont
  {Childs}}\ and\ \bibinfo {author} {\bibfnamefont {Y.}~\bibnamefont {Ge}},\
  }\href {\doibase 10.1103/PhysRevA.89.052337} {\bibfield  {journal} {\bibinfo
  {journal} {Phys. Rev. A}\ }\textbf {\bibinfo {volume} {89}},\ \bibinfo
  {pages} {052337} (\bibinfo {year} {2014})}\BibitemShut {NoStop}%
\bibitem [{\citenamefont {Childs}(2009)}]{Childs2009}%
  \BibitemOpen
  \bibfield  {author} {\bibinfo {author} {\bibfnamefont {A.~M.}\ \bibnamefont
  {Childs}},\ }\href {\doibase 10.1103/PhysRevLett.102.180501} {\bibfield
  {journal} {\bibinfo  {journal} {Phys. Rev. Lett.}\ }\textbf {\bibinfo
  {volume} {102}},\ \bibinfo {pages} {180501} (\bibinfo {year}
  {2009})}\BibitemShut {NoStop}%
\bibitem [{\citenamefont {Childs}\ \emph {et~al.}(2013)\citenamefont {Childs},
  \citenamefont {Gosset},\ and\ \citenamefont {Webb}}]{Childs2013}%
  \BibitemOpen
  \bibfield  {author} {\bibinfo {author} {\bibfnamefont {A.~M.}\ \bibnamefont
  {Childs}}, \bibinfo {author} {\bibfnamefont {D.}~\bibnamefont {Gosset}}, \
  and\ \bibinfo {author} {\bibfnamefont {Z.}~\bibnamefont {Webb}},\ }\href
  {\doibase 10.1126/science.1229957} {\bibfield  {journal} {\bibinfo  {journal}
  {Science}\ }\textbf {\bibinfo {volume} {339}},\ \bibinfo {pages} {791}
  (\bibinfo {year} {2013})}\BibitemShut {NoStop}%
\bibitem [{\citenamefont {Kitagawa}\ \emph {et~al.}(2010)\citenamefont
  {Kitagawa}, \citenamefont {Rudner}, \citenamefont {Berg},\ and\ \citenamefont
  {Demler}}]{PhysRevA.82.033429}%
  \BibitemOpen
  \bibfield  {author} {\bibinfo {author} {\bibfnamefont {T.}~\bibnamefont
  {Kitagawa}}, \bibinfo {author} {\bibfnamefont {M.~S.}\ \bibnamefont
  {Rudner}}, \bibinfo {author} {\bibfnamefont {E.}~\bibnamefont {Berg}}, \ and\
  \bibinfo {author} {\bibfnamefont {E.}~\bibnamefont {Demler}},\ }\href
  {\doibase 10.1103/PhysRevA.82.033429} {\bibfield  {journal} {\bibinfo
  {journal} {Phys. Rev. A}\ }\textbf {\bibinfo {volume} {82}},\ \bibinfo
  {pages} {033429} (\bibinfo {year} {2010})}\BibitemShut {NoStop}%
\bibitem [{\citenamefont {Kitagawa}(2012)}]{Kitagawa2012a}%
  \BibitemOpen
  \bibfield  {author} {\bibinfo {author} {\bibfnamefont {T.}~\bibnamefont
  {Kitagawa}},\ }\href {\doibase 10.1007/s11128-012-0425-4} {\bibfield
  {journal} {\bibinfo  {journal} {Quantum Inf. Process}\ }\textbf {\bibinfo
  {volume} {11}},\ \bibinfo {pages} {1107} (\bibinfo {year}
  {2012})}\BibitemShut {NoStop}%
\bibitem [{\citenamefont {Kitagawa}\ \emph {et~al.}(2012)\citenamefont
  {Kitagawa}, \citenamefont {Broome}, \citenamefont {Fedrizzi}, \citenamefont
  {Rudner}, \citenamefont {Berg}, \citenamefont {Kassal}, \citenamefont
  {Aspuru-Guzik}, \citenamefont {Demler},\ and\ \citenamefont
  {White}}]{Kitagawa2012b}%
  \BibitemOpen
  \bibfield  {author} {\bibinfo {author} {\bibfnamefont {T.}~\bibnamefont
  {Kitagawa}}, \bibinfo {author} {\bibfnamefont {M.~A.}\ \bibnamefont
  {Broome}}, \bibinfo {author} {\bibfnamefont {A.}~\bibnamefont {Fedrizzi}},
  \bibinfo {author} {\bibfnamefont {M.~S.}\ \bibnamefont {Rudner}}, \bibinfo
  {author} {\bibfnamefont {E.}~\bibnamefont {Berg}}, \bibinfo {author}
  {\bibfnamefont {I.}~\bibnamefont {Kassal}}, \bibinfo {author} {\bibfnamefont
  {A.}~\bibnamefont {Aspuru-Guzik}}, \bibinfo {author} {\bibfnamefont
  {E.}~\bibnamefont {Demler}}, \ and\ \bibinfo {author} {\bibfnamefont {A.~G.}\
  \bibnamefont {White}},\ }\href {\doibase 10.1038/ncomms1872} {\bibfield
  {journal} {\bibinfo  {journal} {Nat. Commun.}\ }\textbf {\bibinfo {volume}
  {3}},\ \bibinfo {pages} {882} (\bibinfo {year} {2012})}\BibitemShut {NoStop}%
\bibitem [{\citenamefont {Eisert}\ \emph {et~al.}(2015)\citenamefont {Eisert},
  \citenamefont {Friesdorf},\ and\ \citenamefont {Gogolin}}]{Eisert2015}%
  \BibitemOpen
  \bibfield  {author} {\bibinfo {author} {\bibfnamefont {J.}~\bibnamefont
  {Eisert}}, \bibinfo {author} {\bibfnamefont {M.}~\bibnamefont {Friesdorf}}, \
  and\ \bibinfo {author} {\bibfnamefont {C.}~\bibnamefont {Gogolin}},\ }\href
  {\doibase 10.1038/nphys3215} {\bibfield  {journal} {\bibinfo  {journal} {Nat.
  Phys.}\ }\textbf {\bibinfo {volume} {11}},\ \bibinfo {pages} {124} (\bibinfo
  {year} {2015})}\BibitemShut {NoStop}%
\bibitem [{\citenamefont {Lang}\ \emph {et~al.}(2018)\citenamefont {Lang},
  \citenamefont {Frank},\ and\ \citenamefont
  {Halimeh}}]{PhysRevLett.121.130603}%
  \BibitemOpen
  \bibfield  {author} {\bibinfo {author} {\bibfnamefont {J.}~\bibnamefont
  {Lang}}, \bibinfo {author} {\bibfnamefont {B.}~\bibnamefont {Frank}}, \ and\
  \bibinfo {author} {\bibfnamefont {J.~C.}\ \bibnamefont {Halimeh}},\ }\href
  {\doibase 10.1103/PhysRevLett.121.130603} {\bibfield  {journal} {\bibinfo
  {journal} {Phys. Rev. Lett.}\ }\textbf {\bibinfo {volume} {121}},\ \bibinfo
  {pages} {130603} (\bibinfo {year} {2018})}\BibitemShut {NoStop}%
\bibitem [{\citenamefont {Heyl}(2018)}]{heyl2018dynamical}%
  \BibitemOpen
  \bibfield  {author} {\bibinfo {author} {\bibfnamefont {M.}~\bibnamefont
  {Heyl}},\ }\href {\doibase 10.1088/1361-6633/aaaf9a} {\bibfield  {journal}
  {\bibinfo  {journal} {Rep. Prog. Phys.}\ }\textbf {\bibinfo {volume} {81}},\
  \bibinfo {pages} {054001} (\bibinfo {year} {2018})}\BibitemShut {NoStop}%
\bibitem [{\citenamefont {Xu}\ \emph {et~al.}(2018{\natexlab{a}})\citenamefont
  {Xu}, \citenamefont {Wang}, \citenamefont {Heyl}, \citenamefont {Budich},
  \citenamefont {Pan}, \citenamefont {Chen}, \citenamefont {Jan}, \citenamefont
  {Sun}, \citenamefont {Xu}, \citenamefont {Han} \emph
  {et~al.}}]{xu2018measuring}%
  \BibitemOpen
  \bibfield  {author} {\bibinfo {author} {\bibfnamefont {X.-Y.}\ \bibnamefont
  {Xu}}, \bibinfo {author} {\bibfnamefont {Q.-Q.}\ \bibnamefont {Wang}},
  \bibinfo {author} {\bibfnamefont {M.}~\bibnamefont {Heyl}}, \bibinfo {author}
  {\bibfnamefont {J.~C.}\ \bibnamefont {Budich}}, \bibinfo {author}
  {\bibfnamefont {W.-W.}\ \bibnamefont {Pan}}, \bibinfo {author} {\bibfnamefont
  {Z.}~\bibnamefont {Chen}}, \bibinfo {author} {\bibfnamefont {M.}~\bibnamefont
  {Jan}}, \bibinfo {author} {\bibfnamefont {K.}~\bibnamefont {Sun}}, \bibinfo
  {author} {\bibfnamefont {J.-S.}\ \bibnamefont {Xu}}, \bibinfo {author}
  {\bibfnamefont {Y.-J.}\ \bibnamefont {Han}},  \emph {et~al.},\ }\href@noop {}
  {\bibfield  {journal} {\bibinfo  {journal} {arXiv:1808.03930 [quant-ph]}\ }
  (\bibinfo {year} {2018}{\natexlab{a}})}\BibitemShut {NoStop}%
\bibitem [{\citenamefont {Parrondo}\ \emph {et~al.}(2015)\citenamefont
  {Parrondo}, \citenamefont {Horowitz},\ and\ \citenamefont
  {Sagawa}}]{Parrondo2015}%
  \BibitemOpen
  \bibfield  {author} {\bibinfo {author} {\bibfnamefont {J.~M.~R.}\
  \bibnamefont {Parrondo}}, \bibinfo {author} {\bibfnamefont {J.~M.}\
  \bibnamefont {Horowitz}}, \ and\ \bibinfo {author} {\bibfnamefont
  {T.}~\bibnamefont {Sagawa}},\ }\href {\doibase 10.1038/nphys3230} {\bibfield
  {journal} {\bibinfo  {journal} {Nat. Phys.}\ }\textbf {\bibinfo {volume}
  {11}},\ \bibinfo {pages} {131} (\bibinfo {year} {2015})}\BibitemShut
  {NoStop}%
\bibitem [{\citenamefont {Garnerone}(2012)}]{Garnerone2012}%
  \BibitemOpen
  \bibfield  {author} {\bibinfo {author} {\bibfnamefont {S.}~\bibnamefont
  {Garnerone}},\ }\href {\doibase 10.1103/PhysRevA.86.032342} {\bibfield
  {journal} {\bibinfo  {journal} {Phys. Rev. A}\ }\textbf {\bibinfo {volume}
  {86}},\ \bibinfo {pages} {032342} (\bibinfo {year} {2012})}\BibitemShut
  {NoStop}%
\bibitem [{\citenamefont {Romanelli}(2012)}]{Romanelli2012}%
  \BibitemOpen
  \bibfield  {author} {\bibinfo {author} {\bibfnamefont {A.}~\bibnamefont
  {Romanelli}},\ }\href {\doibase 10.1103/PhysRevA.85.012319} {\bibfield
  {journal} {\bibinfo  {journal} {Phys. Rev. A}\ }\textbf {\bibinfo {volume}
  {85}},\ \bibinfo {pages} {012319} (\bibinfo {year} {2012})}\BibitemShut
  {NoStop}%
\bibitem [{\citenamefont {Romanelli}\ \emph {et~al.}(2014)\citenamefont
  {Romanelli}, \citenamefont {Donangelo}, \citenamefont {Portugal},\ and\
  \citenamefont {Marquezino}}]{Romanelli2014}%
  \BibitemOpen
  \bibfield  {author} {\bibinfo {author} {\bibfnamefont {A.}~\bibnamefont
  {Romanelli}}, \bibinfo {author} {\bibfnamefont {R.}~\bibnamefont
  {Donangelo}}, \bibinfo {author} {\bibfnamefont {R.}~\bibnamefont {Portugal}},
  \ and\ \bibinfo {author} {\bibfnamefont {F.~d.~L.}\ \bibnamefont
  {Marquezino}},\ }\href {\doibase 10.1103/PhysRevA.90.022329} {\bibfield
  {journal} {\bibinfo  {journal} {Phys. Rev. A}\ }\textbf {\bibinfo {volume}
  {90}},\ \bibinfo {pages} {022329} (\bibinfo {year} {2014})}\BibitemShut
  {NoStop}%
\bibitem [{\citenamefont {Romanelli}(2015)}]{Romanelli2015}%
  \BibitemOpen
  \bibfield  {author} {\bibinfo {author} {\bibfnamefont {A.}~\bibnamefont
  {Romanelli}},\ }\href {\doibase https://doi.org/10.1016/j.physa.2015.03.084}
  {\bibfield  {journal} {\bibinfo  {journal} {Physica A}\ }\textbf {\bibinfo
  {volume} {434}},\ \bibinfo {pages} {111} (\bibinfo {year}
  {2015})}\BibitemShut {NoStop}%
\bibitem [{\citenamefont {Flitney}\ \emph {et~al.}(2004)\citenamefont
  {Flitney}, \citenamefont {Abbott},\ and\ \citenamefont
  {Johnson}}]{Flitney2004}%
  \BibitemOpen
  \bibfield  {author} {\bibinfo {author} {\bibfnamefont {A.~P.}\ \bibnamefont
  {Flitney}}, \bibinfo {author} {\bibfnamefont {D.}~\bibnamefont {Abbott}}, \
  and\ \bibinfo {author} {\bibfnamefont {N.~F.}\ \bibnamefont {Johnson}},\
  }\href {http://stacks.iop.org/0305-4470/37/i=30/a=013} {\bibfield  {journal}
  {\bibinfo  {journal} {J. Phys. A: Math. Gen.}\ }\textbf {\bibinfo {volume}
  {37}},\ \bibinfo {pages} {7581} (\bibinfo {year} {2004})}\BibitemShut
  {NoStop}%
\bibitem [{\citenamefont {Chandrashekar}\ and\ \citenamefont
  {Banerjee}(2011)}]{chandrashekar2011parrondo}%
  \BibitemOpen
  \bibfield  {author} {\bibinfo {author} {\bibfnamefont {C.}~\bibnamefont
  {Chandrashekar}}\ and\ \bibinfo {author} {\bibfnamefont {S.}~\bibnamefont
  {Banerjee}},\ }\href {\doibase 10.1016/j.physleta.2011.02.071} {\bibfield
  {journal} {\bibinfo  {journal} {Phys. Lett. A}\ }\textbf {\bibinfo {volume}
  {375}},\ \bibinfo {pages} {1553} (\bibinfo {year} {2011})}\BibitemShut
  {NoStop}%
\bibitem [{\citenamefont {Flitney}(2012)}]{flitney2012quantum}%
  \BibitemOpen
  \bibfield  {author} {\bibinfo {author} {\bibfnamefont {A.~P.}\ \bibnamefont
  {Flitney}},\ }\href@noop {} {\bibfield  {journal} {\bibinfo  {journal}
  {arXiv:1209.2252 [quant-ph]}\ } (\bibinfo {year} {2012})}\BibitemShut
  {NoStop}%
\bibitem [{\citenamefont {Li}\ \emph {et~al.}(2013)\citenamefont {Li},
  \citenamefont {Zhang},\ and\ \citenamefont {Guo}}]{li2013quantum}%
  \BibitemOpen
  \bibfield  {author} {\bibinfo {author} {\bibfnamefont {M.}~\bibnamefont
  {Li}}, \bibinfo {author} {\bibfnamefont {Y.-S.}\ \bibnamefont {Zhang}}, \
  and\ \bibinfo {author} {\bibfnamefont {G.-C.}\ \bibnamefont {Guo}},\ }\href
  {\doibase 10.1142/S0219477502000920} {\bibfield  {journal} {\bibinfo
  {journal} {Fluctuation Noise Lett.}\ }\textbf {\bibinfo {volume} {12}},\
  \bibinfo {pages} {1350024} (\bibinfo {year} {2013})}\BibitemShut {NoStop}%
\bibitem [{\citenamefont {Rajendran}\ and\ \citenamefont
  {Benjamin}(2018{\natexlab{a}})}]{rajendran2018playing}%
  \BibitemOpen
  \bibfield  {author} {\bibinfo {author} {\bibfnamefont {J.}~\bibnamefont
  {Rajendran}}\ and\ \bibinfo {author} {\bibfnamefont {C.}~\bibnamefont
  {Benjamin}},\ }\href {\doibase 10.1209/0295-5075/122/40004} {\bibfield
  {journal} {\bibinfo  {journal} {EPL}\ }\textbf {\bibinfo {volume} {122}},\
  \bibinfo {pages} {40004} (\bibinfo {year} {2018}{\natexlab{a}})}\BibitemShut
  {NoStop}%
\bibitem [{\citenamefont {Rajendran}\ and\ \citenamefont
  {Benjamin}(2018{\natexlab{b}})}]{rajendran2018implementing}%
  \BibitemOpen
  \bibfield  {author} {\bibinfo {author} {\bibfnamefont {J.}~\bibnamefont
  {Rajendran}}\ and\ \bibinfo {author} {\bibfnamefont {C.}~\bibnamefont
  {Benjamin}},\ }\href {\doibase 10.1098/rsos.171599} {\bibfield  {journal}
  {\bibinfo  {journal} {Royal Soc. Open Sci.}\ }\textbf {\bibinfo {volume}
  {5}},\ \bibinfo {pages} {171599} (\bibinfo {year}
  {2018}{\natexlab{b}})}\BibitemShut {NoStop}%
\bibitem [{\citenamefont {Pires}\ and\ \citenamefont
  {Queir\'os}(2020)}]{Pires2020}%
  \BibitemOpen
  \bibfield  {author} {\bibinfo {author} {\bibfnamefont {M.~A.}\ \bibnamefont
  {Pires}}\ and\ \bibinfo {author} {\bibfnamefont {S.~M.~D.}\ \bibnamefont
  {Queir\'os}},\ }\href {\doibase 10.1103/PhysRevE.102.042124} {\bibfield
  {journal} {\bibinfo  {journal} {Phys. Rev. E}\ }\textbf {\bibinfo {volume}
  {102}},\ \bibinfo {pages} {042124} (\bibinfo {year} {2020})}\BibitemShut
  {NoStop}%
\bibitem [{\citenamefont {Walczak}\ and\ \citenamefont
  {Bauer}(2021)}]{Walczak2021}%
  \BibitemOpen
  \bibfield  {author} {\bibinfo {author} {\bibfnamefont {Z.}~\bibnamefont
  {Walczak}}\ and\ \bibinfo {author} {\bibfnamefont {J.~H.}\ \bibnamefont
  {Bauer}},\ }\href {\doibase 10.1103/PhysRevE.104.064209} {\bibfield
  {journal} {\bibinfo  {journal} {Phys. Rev. E}\ }\textbf {\bibinfo {volume}
  {104}},\ \bibinfo {pages} {064209} (\bibinfo {year} {2021})}\BibitemShut
  {NoStop}%
\bibitem [{\citenamefont {Schreiber}\ \emph {et~al.}(2011)\citenamefont
  {Schreiber}, \citenamefont {Cassemiro}, \citenamefont
  {Poto\ifmmode~\check{c}\else \v{c}\fi{}ek}, \citenamefont {G\'abris},
  \citenamefont {Jex},\ and\ \citenamefont
  {Silberhorn}}]{PhysRevLett.106.180403}%
  \BibitemOpen
  \bibfield  {author} {\bibinfo {author} {\bibfnamefont {A.}~\bibnamefont
  {Schreiber}}, \bibinfo {author} {\bibfnamefont {K.~N.}\ \bibnamefont
  {Cassemiro}}, \bibinfo {author} {\bibfnamefont {V.}~\bibnamefont
  {Poto\ifmmode~\check{c}\else \v{c}\fi{}ek}}, \bibinfo {author} {\bibfnamefont
  {A.}~\bibnamefont {G\'abris}}, \bibinfo {author} {\bibfnamefont
  {I.}~\bibnamefont {Jex}}, \ and\ \bibinfo {author} {\bibfnamefont
  {C.}~\bibnamefont {Silberhorn}},\ }\href {\doibase
  10.1103/PhysRevLett.106.180403} {\bibfield  {journal} {\bibinfo  {journal}
  {Phys. Rev. Lett.}\ }\textbf {\bibinfo {volume} {106}},\ \bibinfo {pages}
  {180403} (\bibinfo {year} {2011})}\BibitemShut {NoStop}%
\bibitem [{\citenamefont {Kumar}\ \emph {et~al.}(2018)\citenamefont {Kumar},
  \citenamefont {Balu}, \citenamefont {Laflamme},\ and\ \citenamefont
  {Chandrashekar}}]{PhysRevA.97.012116}%
  \BibitemOpen
  \bibfield  {author} {\bibinfo {author} {\bibfnamefont {N.~P.}\ \bibnamefont
  {Kumar}}, \bibinfo {author} {\bibfnamefont {R.}~\bibnamefont {Balu}},
  \bibinfo {author} {\bibfnamefont {R.}~\bibnamefont {Laflamme}}, \ and\
  \bibinfo {author} {\bibfnamefont {C.~M.}\ \bibnamefont {Chandrashekar}},\
  }\href {\doibase 10.1103/PhysRevA.97.012116} {\bibfield  {journal} {\bibinfo
  {journal} {Phys. Rev. A}\ }\textbf {\bibinfo {volume} {97}},\ \bibinfo
  {pages} {012116} (\bibinfo {year} {2018})}\BibitemShut {NoStop}%
\bibitem [{\citenamefont {Vieira}\ \emph {et~al.}(2013)\citenamefont {Vieira},
  \citenamefont {Amorim},\ and\ \citenamefont
  {Rigolin}}]{PhysRevLett.111.180503}%
  \BibitemOpen
  \bibfield  {author} {\bibinfo {author} {\bibfnamefont {R.}~\bibnamefont
  {Vieira}}, \bibinfo {author} {\bibfnamefont {E.~P.~M.}\ \bibnamefont
  {Amorim}}, \ and\ \bibinfo {author} {\bibfnamefont {G.}~\bibnamefont
  {Rigolin}},\ }\href {\doibase 10.1103/PhysRevLett.111.180503} {\bibfield
  {journal} {\bibinfo  {journal} {Phys. Rev. Lett.}\ }\textbf {\bibinfo
  {volume} {111}},\ \bibinfo {pages} {180503} (\bibinfo {year}
  {2013})}\BibitemShut {NoStop}%
\bibitem [{\citenamefont {Vieira}\ \emph {et~al.}(2014)\citenamefont {Vieira},
  \citenamefont {Amorim},\ and\ \citenamefont {Rigolin}}]{Vieira2014}%
  \BibitemOpen
  \bibfield  {author} {\bibinfo {author} {\bibfnamefont {R.}~\bibnamefont
  {Vieira}}, \bibinfo {author} {\bibfnamefont {E.~P.~M.}\ \bibnamefont
  {Amorim}}, \ and\ \bibinfo {author} {\bibfnamefont {G.}~\bibnamefont
  {Rigolin}},\ }\href {\doibase 10.1103/PhysRevA.89.042307} {\bibfield
  {journal} {\bibinfo  {journal} {Phys. Rev. A}\ }\textbf {\bibinfo {volume}
  {89}},\ \bibinfo {pages} {042307} (\bibinfo {year} {2014})}\BibitemShut
  {NoStop}%
\bibitem [{\citenamefont {Wang}\ \emph {et~al.}(2018)\citenamefont {Wang},
  \citenamefont {Xu}, \citenamefont {Pan}, \citenamefont {Sun}, \citenamefont
  {Xu}, \citenamefont {Chen}, \citenamefont {Han}, \citenamefont {Li},\ and\
  \citenamefont {Guo}}]{wang2018dynamic}%
  \BibitemOpen
  \bibfield  {author} {\bibinfo {author} {\bibfnamefont {Q.-Q.}\ \bibnamefont
  {Wang}}, \bibinfo {author} {\bibfnamefont {X.-Y.}\ \bibnamefont {Xu}},
  \bibinfo {author} {\bibfnamefont {W.-W.}\ \bibnamefont {Pan}}, \bibinfo
  {author} {\bibfnamefont {K.}~\bibnamefont {Sun}}, \bibinfo {author}
  {\bibfnamefont {J.-S.}\ \bibnamefont {Xu}}, \bibinfo {author} {\bibfnamefont
  {G.}~\bibnamefont {Chen}}, \bibinfo {author} {\bibfnamefont {Y.-J.}\
  \bibnamefont {Han}}, \bibinfo {author} {\bibfnamefont {C.-F.}\ \bibnamefont
  {Li}}, \ and\ \bibinfo {author} {\bibfnamefont {G.-C.}\ \bibnamefont {Guo}},\
  }\href {\doibase 10.1364/OPTICA.5.001136} {\bibfield  {journal} {\bibinfo
  {journal} {Optica}\ }\textbf {\bibinfo {volume} {5}},\ \bibinfo {pages}
  {1136} (\bibinfo {year} {2018})}\BibitemShut {NoStop}%
\bibitem [{\citenamefont {Jan}\ \emph {et~al.}(2020)\citenamefont {Jan},
  \citenamefont {Wang}, \citenamefont {Xu}, \citenamefont {Pan}, \citenamefont
  {Chen}, \citenamefont {Han}, \citenamefont {Li}, \citenamefont {Guo},\ and\
  \citenamefont {Abbott}}]{jan2020}%
  \BibitemOpen
  \bibfield  {author} {\bibinfo {author} {\bibfnamefont {M.}~\bibnamefont
  {Jan}}, \bibinfo {author} {\bibfnamefont {Q.-Q.}\ \bibnamefont {Wang}},
  \bibinfo {author} {\bibfnamefont {X.-Y.}\ \bibnamefont {Xu}}, \bibinfo
  {author} {\bibfnamefont {W.-W.}\ \bibnamefont {Pan}}, \bibinfo {author}
  {\bibfnamefont {Z.}~\bibnamefont {Chen}}, \bibinfo {author} {\bibfnamefont
  {Y.-J.}\ \bibnamefont {Han}}, \bibinfo {author} {\bibfnamefont {C.-F.}\
  \bibnamefont {Li}}, \bibinfo {author} {\bibfnamefont {G.-C.}\ \bibnamefont
  {Guo}}, \ and\ \bibinfo {author} {\bibfnamefont {D.}~\bibnamefont {Abbott}},\
  }\href {\doibase 10.1002/qute.201900127} {\bibfield  {journal} {\bibinfo
  {journal} {Adv. Quantum Technol.}\ }\textbf {\bibinfo {volume} {3}},\
  \bibinfo {pages} {1900127} (\bibinfo {year} {2020})}\BibitemShut {NoStop}%
\bibitem [{\citenamefont {Xu}\ \emph {et~al.}(2018{\natexlab{b}})\citenamefont
  {Xu}, \citenamefont {Wang}, \citenamefont {Pan}, \citenamefont {Sun},
  \citenamefont {Xu}, \citenamefont {Chen}, \citenamefont {Tang}, \citenamefont
  {Gong}, \citenamefont {Han}, \citenamefont {Li},\ and\ \citenamefont
  {Guo}}]{Xu2018}%
  \BibitemOpen
  \bibfield  {author} {\bibinfo {author} {\bibfnamefont {X.-Y.}\ \bibnamefont
  {Xu}}, \bibinfo {author} {\bibfnamefont {Q.-Q.}\ \bibnamefont {Wang}},
  \bibinfo {author} {\bibfnamefont {W.-W.}\ \bibnamefont {Pan}}, \bibinfo
  {author} {\bibfnamefont {K.}~\bibnamefont {Sun}}, \bibinfo {author}
  {\bibfnamefont {J.-S.}\ \bibnamefont {Xu}}, \bibinfo {author} {\bibfnamefont
  {G.}~\bibnamefont {Chen}}, \bibinfo {author} {\bibfnamefont {J.-S.}\
  \bibnamefont {Tang}}, \bibinfo {author} {\bibfnamefont {M.}~\bibnamefont
  {Gong}}, \bibinfo {author} {\bibfnamefont {Y.-J.}\ \bibnamefont {Han}},
  \bibinfo {author} {\bibfnamefont {C.-F.}\ \bibnamefont {Li}}, \ and\ \bibinfo
  {author} {\bibfnamefont {G.-C.}\ \bibnamefont {Guo}},\ }\href {\doibase
  10.1103/PhysRevLett.120.260501} {\bibfield  {journal} {\bibinfo  {journal}
  {Phys. Rev. Lett.}\ }\textbf {\bibinfo {volume} {120}},\ \bibinfo {pages}
  {260501} (\bibinfo {year} {2018}{\natexlab{b}})}\BibitemShut {NoStop}%
\end{thebibliography}
\end{document}